\documentstyle[graphics,psfig]{article}
\begin{document}
\textheight=18.5cm
\headheight=0pt
\headsep=0pt
\topmargin=0.5in
\footheight=-3.5in
\textwidth=15cm
\pagenumbering{arabic}
\newcommand{\Ha}	{H$\alpha$}
\newcommand{\Msun}	{\hb{$\rm M_{\odot}$}}
\newcommand{\HII}	{H$\,${\sc ii}}

\title{{\bf Statistical and Energetic Constraints in Population Synthesis Models}
\author{Alberto Buzzoni\\
 Osservatorio Astronomico di Brera\\
 Via Bianchi, 46 - 23807 Merate (Lc), Italy}}
\maketitle

\begin{abstract}
Physical and numerical constraints in building up self-consistent population 
synthesis models are briefly analysed discussing their application to most of 
the current synthesis codes widely adopted in Galactic and extragalactic 
studies.

Major focus is given in particular to the numerical effects of discrete 
isochrone sampling as well as to the energetic constraints for a proper 
normalization of the main sequence vs. post-MS luminosity contribution in 
simple stellar populations.

The case of template models for present-day elliptical galaxies is reviewed
in more detail comparing the results from different synthesis codes.
\end{abstract}

\section{Introduction}

Stellar population synthesis has extensively been used in the recent 
years as a main theoretical tool for Galactic and extragalactic studies.
In particular, {\it evolutionary} synthesis (specifically relying on 
theoretical tracks for stellar evolution) has quickly become the most popular 
technique to investigate spectrophotometric properties of external galaxies
(cf. e.g. Ellis 1997).

While, in its essence, a synthesis procedure is a quite simple task-- we
just need to sum up luminosity from stars along isochrone bins to 
obtain the luminosity of the population as a whole in the different 
photometric bands-- there are however a number of crucial constraints 
which we must pay attention to in order to achieve a {\it physically} 
(and not only {\it numerically}) self-consistent output model.

In this paper we would briefly account for some of these features discussing
some reference cases from the main synthesis codes widely adopted in the 
recent literature.

It is worth stressing that the main aim of our analysis is {\it not} to
single out any {\it superior} model or trying any ``ranking'' among the
available codes, but rather to lead the reader to a more critical appraisal
of the current theoretical framework. This, as a little step toward a better 
use of population synthesis in the interpretative analysis of real stellar 
systems.

\section{Numerical discreteness and statistical sampling}

A basic step when assembling a synthetic stellar population concerns the
way we choose to sample the theoretical isochrone.
Focussing our attention to the general case of a simple stellar population 
(SSP) intended as a single generation of coeval stars with fixed distinctive 
parameters and mass distribution, we can envisage two possible procedures 
in this regard:

{\it i)} we might decide to add stars discretely by random sampling according to
the isochrone locus and count-density map or

{\it ii)} we might attribute each luminosity bin a weight according to its 
expected relative density normalizing to the total size of the population.

An example of case {\it (i)} is shown in Fig. 1 for a 15 Gyr Salpeter SSP 
with solar metallicity according to Buzzoni's (1989) code.
The advantage of this approach is to provide a more vivid representation
of the synthetic c-m diagram of a stellar population, giving a direct hint of 
the relative distribution of the stars along the different evolutionary phases.

On the other hand, as a main drawback, we might severely be affected by 
normalization problems and statistical undersampling of all those 
fast-evolving phases if they predict a few or even less than one star in the 
whole sample.

For example, according to the Renzini and Buzzoni (1986) fuel consumption
theorem (FCT), about 15 stars should be expected along the asymptotic giant 
branch (AGB) in the SSP of Fig. 1. 
This means that in general a $1/\sqrt 15 \sim 25\%$ statistical uncertainty 
in the AGB luminosity contribution could affect a random simulation 
of this size. While this is a negligible detail when observing in the $V$ band,
things might be very different in the infrared where AGB stars provide a much 
larger fraction of the total luminosity of the population.
Following Buzzoni's (1993) calculations, in the case of the synthetic c-m of 
Fig. 1 a statistical fluctuation as large as $\pm 0.2$ mag should eventually 
be expected in the integrated $V-K$ color.  Things would get even worst as 
far as we move to a younger age because of a faster post main sequence 
(post-MS) evolution.

Note that while statistical uncertainty from sample discreteness could be 
a {\it real} (and often unavoidable) case when observing for instance Galactic 
globular clusters, the effect could in principle be overcome in theoretical 
simulations but at cost of increasing the random sample to an exceedingly 
larger size ($\geq 10^7$ stars).

A synthesis procedure that relies on case {\it (ii)} is on the contrary
more suitable to achieve an unbiased estimate of the SSP integrated properties.
In the latter case in facts the contribution from each isochrone bin is taken
{\it at its face value}, and correctly accounted for according to the 
theoretical luminosity function. Integrated magnitudes and colors will 
therefore not depend on the size of the population.

As a partial limitation, the latter method somewhat misses a direct picture 
of star distribution merely restraining the SSP c-m diagram to the locus of the 
theoretical isochrone. Compare for example the right-hand panel of Fig. 2 
displaying the equivalent SSP of Fig. 1 as it appears from the Padova 
synthesis code (Bertelli {\it et al.} 1994).

\begin{figure}
\centerline{\psfig{bbllx=68pt,bblly=181pt,bburx=513pt,bbury=613pt,file=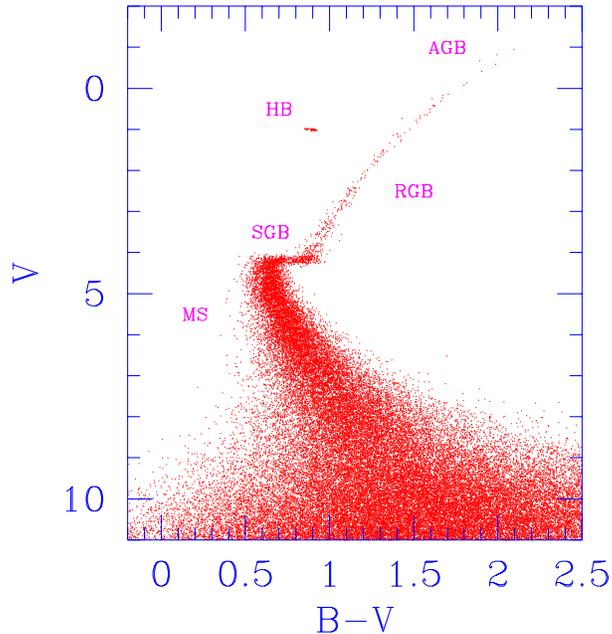,angle=0,width=9cm,clip=}}
\caption{Synthetic c-m diagram for a 15 Gyr Salpeter SSP of solar metallicity
(from Buzzoni's 1989 code). The random simulation consists of $10^5$ stars, a 
size comparable with a typical Galactic globular cluster. Data have been added 
a Poissonian photometric error such as $\sigma V = 0.03$ mag at $V = +4$. The 
main stellar evolutionary phases are labelled in the figure. Note the effect 
of sample discreteness in the fast-evolving phases (like for instance near 
the RGB tip or along the AGB phase) that result severely undersampled.}
\end{figure}

\begin{figure}
\centerline{\psfig{bbllx=65pt,bblly=218pt,bburx=554pt,bbury=477pt,file=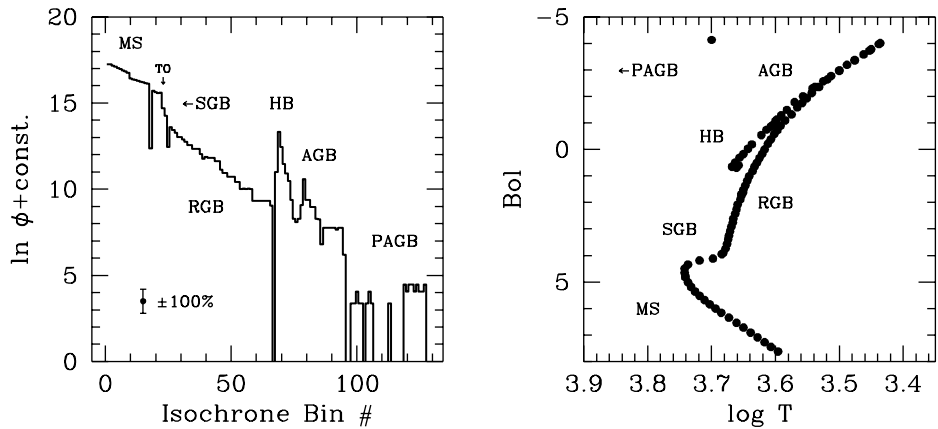,angle=0,width=13cm,clip=}}
\caption{Theoretical luminosity function {\it (left panel)} and isochrone locus
{\it (right panel)} for a 15 Gyr Salpeter SSP of solar metallicity from the 
Padova model database (Bertelli {\it et al.} 1994). The main evolutionary 
phases are labelled in the figure. See text for discussion.}
\end{figure}

Preliminary to any fair determination of SSP properties however (either via
case {\it (i)} or {\it (ii)}), a more general problem that deserves our
attention for its potential impact on the synthesis results deals with the 
numerical representation of the star number counts along the theoretical 
isochrone.

In a general approach to isochrone calculation relying on the interpolation
within a grid of stellar tracks, the number of stars in each bin 
eventually derives from the adopted IMF and the estimated width in stellar mass
of the ``j-th'' cell ($\delta M_j$), namely $N_j = IMF \times 
(\delta M)_j$. In spite of any smooth IMF, we should rather expect a 
``roughness'' in the bin number counts such as $\sigma N \sim \sigma 
(\delta M)$. A large uncertainty in the values of $(\delta M)_j$ could in facts
easily derive from numerical fluctuations in the interpolation algorithm given
the extremely thin range of mass variation along post-MS isochrone bins.
For example, a red giant branch (RGB) in an old SSP sampled with 50 bins
will typically imply $\delta M \sim 0.0006 M_\odot$ (cf. next section).
This means that a $10^{-4}$ relative accuracy in the mass interpolation
algorithm should be achieved in order to assure a 10\% accuracy in the 
number counts along the theoretical isochrone.

This effect is often neglected in the model analysis but the left-hand panel
of Fig. 2 cleary shows how pervasive it can be as far as the SSP luminosity
function is taken into account in more detail. Note for example in the
figure the two ``glitches'' about the TO region and the discontinuity
about the RGB tip (about bin sequence numbers 50-65) as well as the 
scatter along the post-AGB evolution (beyond bin number 100).
Again, one should carefully consider the differential impact of this effect 
when accounting for the SSP photometric properties in different wavelength 
ranges.

The problem of a fair determination of the isochrone luminosity function 
eventually calls for another primary question in evolutionary population 
synthesis dealing with a physical match of MS and post-MS evolution 
in a SSP, as we will discuss in the next section.

\section{Energetic constraints: Post-MS vs. MS contribution}

As a general feature, SSP post-MS evolution is recognized to always proceed 
much faster than stellar MS lifetime at every age.
If we get an estimate of the rate of change of the MS stellar Turn Off (TO)
mass ($M_{TO}$), then the expected range ($\Delta M_*$) of actual stellar 
masses that are experiencing post-MS evolution over a lifetime $\tau_{PMS}$ 
at a given SSP age ($T$) is
$$ \Delta M_* = {dM_{TO}\over d\; t} ~~ \tau_{PMS} ~~ \sim ~~ {M_{TO}\over T} ~~ \tau_{PMS}. \eqno (1) $$
\noindent
This means, in other words, that 
$${\Delta M_* \over M_{TO}}~~ \sim ~~ {\tau_{PMS} \over T}~~ \leq ~~ 0.1. \eqno (2) $$
As a consequence, if stars of $1\ M_\odot$ are crossing the TO point in an
old SSP then stars of just $1.1 M_\odot$ are exhausting their post-MS evolution
becoming white dwarfs. It is clear therefore that the SSP post-MS isochrone
nearly merges, to a first aproximation, with the evolutionary track of
a star of fixed mass $M_* \to M_{TO}$.
In this sense, post-MS lifetime from the stellar track can easily be converted
into relative star number counts per isochrone bin:

$$N_j/N_{PMS} = \tau_j/\tau_{PMS} \eqno (3)$$

where $\tau_j$ is the lifetime across the j-th bin, so that $\sum_{PMS} \tau_j = \tau_{PMS}$
and $\sum_{PMS} N_j = N_{PMS}$. Note in this regard a substantially difference
(and a much better performance) of this approach to the definition of the 
isochrone luminosity function with respect to the classical interpolation 
algorithm discussed in previous section. With this method in facts we could
fully overcome any numerical instability in the bin star counts as $\tau_j$
is a natural output of the stellar evolutionary track.

As far as the MS evolution is concerned, on the contrary, MS number counts 
simply derives from the IMF so that for a Salpeter power law we have
$N_j \propto M_j^{-s}$ per bin of fixed width in stellar mass.

The real problem, when assembling the whole synthetic SSP is therefore to 
properly scale MS and post-MS contribution in order to preserve overall 
energetic self-consistency in the model.

Clearly, a simple match by grafting contiguous isochrone bins 
at the MS top and at the Post-MS bottom  {\it cannot be a safe solution} 
in this regard as any  uncertainty and numerical noise in the evaluation 
of both extrema would accordingly reverberate into a 
magnified scatter in the contribution of each SSP building block.

Rather than relying on such a {\it differential} normalization procedure, 
one could take advantage of an {\it integral} approach featuring SSP global 
properties via the FCT. In this case, following Renzini and Buzzoni (1986, 
their eq. 14) we have that
$$ {{L_{PMS}}\over{L_{tot}}} = {\cal B} \times {\rm Fuel} = (1.76 \pm 0.4)~m_H. \eqno (4) $$

{\it This simply relates SSP Post-MS luminosity in bolometric to the total 
amount of fuel spent in post-MS evolution by stars of mass $M_{TO}$.}

\begin{figure}
\centerline{\psfig{bbllx=88pt,bblly=227pt,bburx=413pt,bbury=547pt,file=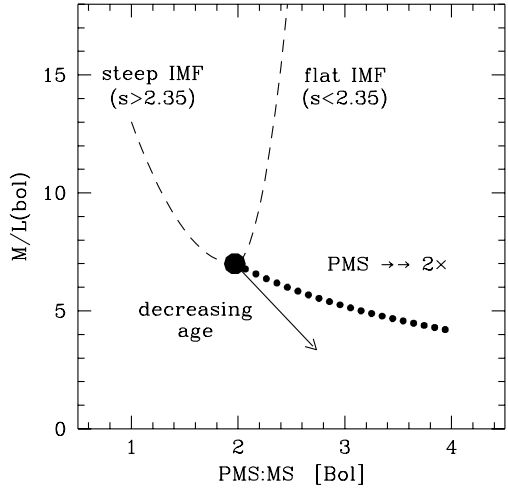,angle=0,width=7cm,clip=}}
\caption{The ``fuel consumption'' diagnostic plot. Bolometric $M/L$ ratio
vs. $L_{PMS}/L_{MS}$ relative contribution is schematically displayed taking
a 15 Gyr Salpeter SSP of solar metallicity as a reference (big dot).
The effect of changing either age or IMF slope is reported on the plot. The 
dotted line tracks the model change by artificially increasing post-MS 
normalization up to a factor of two.}
\end{figure}

In the equation, the normalization factor $\cal B$ is the so-called ``specific 
evolutionary flux''; it turns to be about ${\cal B} = 1.7\pm 0.4~10^{-11}$ [L$_\odot^{-1}$ yr$^{-1}$]
(Buzzoni 1989). The term $m_H$ is the exhausted fuel expressed in Hydrogen-equivalent solar masses.
This quantity is a straightforward output of stellar evolution theory.

One can envisage a simple and direct relationship between the SSP post-MS 
relative contribution from eq. (4) (or its equivalent form in terms of 
$L_{PMS}/L_{MS}$ ratio) and the $M/L$ ratio of the population.
A plot like that in Fig. 3 could provide in this sense an effective tool to
probe energetic self-consistency in synthesis models.
In the figure we took as a reference the standard case of a 15 Gyr Salpeter SSP 
with solar metallicity displaying both the effect of a change in age and
in the IMF slope. Quite interestingly, note that a Salpeter IMF is that 
providing the highest luminosity per unit 
stellar mass in a SSP (Buzzoni 1995). In a flatter IMF in facts most of the
SSP mass belongs to short-living high-mass stars (so that the $M/L$ ratio is
higher than the Salpeter case); for opposite reasons, a steeper IMF will have
a large amount of SSP mass locked into low-luminosity dwarf stars (again
providing a larger $M/L$ ratio with respect to a Salpeter IMF).

\begin{figure}
\centerline{\psfig{bbllx=42pt,bblly=156pt,bburx=529pt,bbury=619pt,file=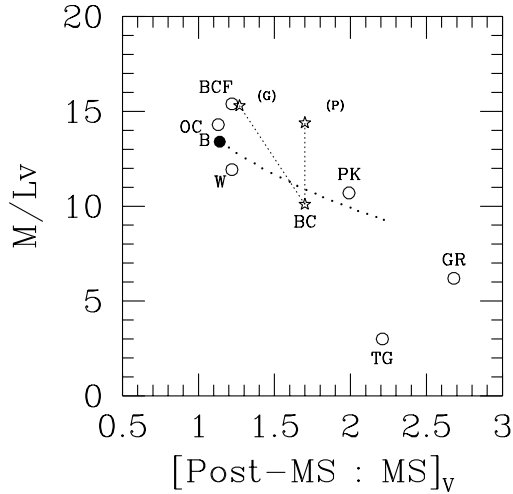,angle=0,width=7cm,clip=}}
\caption{Comparison of template model ellipticals from different theoretical 
codes for population synthesis. Displayed are the predicted $M/L_V$ ratio 
and $L_{PMS}/L_{MS}$ $V$-luminosity partition from the models by 
O'Connell (1976) [OC], Tinsley and Gunn (1976) [TG], Pickles (1985) [PK], 
Guiderdoni and Rocca-Volmerange (1987) [GR], Bruzual and Charlot (1993) [BC], 
Worthey (1994) [W], Bressan {\it et al.} (1994) [BCF], and Buzzoni (1995) [B].
The Bruzual and Charlot (1996) updated code is also reported by matching
both the Padova [(P)] and Geneva [(G)] stellar track database.
The effect of artificially doubling post-MS contribution (then relaxing FCT 
prescriptions) is displayed for Buzzoni's model by the dotted line.}
\end{figure}

The dotted curve in Fig. 3 tracks the locus of a SSP by relaxing the
energetic constraint of the FCT and artificially renormalizing 
the post-MS contribution by increasing luminosity up to a factor of two.
As expected, by overweighting post-MS we will decrease the SSP $M/L$ ratio
(because we are mainly enhancing the total luminosity of the population without 
providing an equivalent extra mass), and obviously move the reference model
toward a higher $L_{PMS}/L_{MS}$ ratio.

An application of this diagnostic plot to the relevant case of template
SSPs from different synthesis codes fitting the present-day elliptical galaxies
is attempted in Fig. 4 updating Buzzoni's (1995) original calculations.
A striking evidence from the figure is that model ellipticals by
Tinsley and Gunn (1976), Pickles (1985), Guiderdoni and Rocca-Volmerange (1987),
and Bruzual and Charlot (1993) do not fully comply with the SSP energetic 
constraint in the sense that all of them appear to be sensibly overestimating 
post-MS contribution.

This discrepancy appears to be partially alleviated in the updated code
of Bruzual and Charlot (1996) (see also Bruzual 1996) once accounting for the 
Geneva set of stellar evolutionary tracks (Schaller {\it et al.} 1992; 
Charbonnel {\it el al.} 1996). It is however at least surprising to note 
that a similar match with the Padova isochrones (Bertelli {\it et al.} 1994) 
leads to a somewhat different model comparing with Bressan's {\it et al.} 
(1994) original calculations based on the same theoretical framework.
In order to recover this discrepancy one should admit that an important 
fraction of the (low?) MS luminosity is missing in the Bruzual and Charlot 
(1996) model.


\small
\begin{thebibliography}{99}
\small
\bibitem[1]{}
Bertelli, G., Bressan, A., Chiosi, C., Fagotto, F., Nasi, E. 1994, {\it A\&ApS}, {\bf 106}, 275
\bibitem[2]{}
Bressan, A., Chiosi, C., and Fagotto, F. 1994, {\it ApJS}, {\bf 94}, 63
\bibitem[3]{}
Bruzual, G. 1996 in ``From Stars to Galaxies'', ASP Conf. Ser., Vol. 98, eds. C. Leitherer, U. Fritze-v. Alvenslebel and J. Huchra (ASP: San Francisco) p. 14
\bibitem[4]{}
Bruzual, G., and Charlot, S. 1993, {\it ApJ}, {\bf 405}, 538
\bibitem[5]{}
Bruzual, G., and Charlot, S. 1996, private communication
\bibitem[6]{}
Buzzoni, A. 1989, {\it ApJS}, {\bf 71}, 817
\bibitem[7]{}
Buzzoni, A. 1993, {\it A\&Ap}, {\bf 275}, 433
\bibitem[8]{}
Buzzoni, A. 1995, {\it ApJS}, {\bf 98}, 69
\bibitem[9]{}
Charbonnel, C., Meynet, G., Maeder, A., Schaerer, D. 1996, {\it A\&ApS}, {\bf 115}, 339
\bibitem[10]{}
Ellis, R.S. 1997, {\it ARAA}, {\bf 35}, 389
\bibitem[11]{}
Guiderdoni, B., and Rocca-Volmerange, B. 1987, {\it A\&A}, {\bf 186}, 1
\bibitem[12]{}
O'Connell, R.W. 1976, {\it ApJ}, {\bf 206}, 370
\bibitem[13]{}
Pickles, A.J. 1985, {\it ApJ}, {\bf 296}, 340
\bibitem[14]{}
Renzini, A., and Buzzoni, A. 1986 in ``Spectral Evolution of Galaxies'', eds. C. Chiosi and A. Renzini (Dordrecht: Reidel) p. 195
\bibitem[15]{}
Schaller, G., Schaerer, D., Meynet, G., Maeder, A. 1992, {\it A\&ApS}, {\bf 96}, 269
\bibitem[16]{}
Tinsley, B.M., and Gunn, J.E. 1976, {\it ApJ}, {\bf 203}, 52
\bibitem[17]{}
Worthey, G. 1994, {\it ApJS}, {\bf 95}, 107
\end{thebibliography}
\end{document}